\documentclass[twocolumn,10pt]{asme2e}

\usepackage{amsmath}
\usepackage{graphicx}
\usepackage{xurl}
\usepackage{orcidlink}
\usepackage{soul}
\usepackage[capitalise]{cleveref}
\usepackage{subcaption}
\usepackage{xcolor}
\usepackage{etoolbox}

\definecolor{preprintbg}{rgb}{0.9,0.9,1}     % exact fill of the Riso box (= blue!10)
\definecolor{preprintlink}{rgb}{0,0,0.5}     % navy, as their links

\newcommand{\aamhandle}{https://hdl.handle.net/10016/51139}     % <-- repository handle
\newcommand{\vorlink}{https://doi.org/10.1115/SSDM2026-183218}  % ASME version of record

\newcommand{\preprintbox}{%
  \noindent{\setlength{\fboxsep}{3pt}%
  \colorbox{preprintbg}{\parbox{\dimexpr\textwidth-2\fboxsep\relax}{%
    \raggedright\small
    {\sffamily\bfseries The AAM is available in Green Open Access at:
    {\color{preprintlink}\expandafter\url\expandafter{\aamhandle}}\\This is the preprint version of the following article:}\\
    Dessena, G., and Pontillo, A., ``Modal Parameter Extraction via
    Propeller-Driven Vibration Testing,'' \emph{Proceedings of the ASME 2026
    Aerospace Structures, Structural Dynamics, and Materials Conference
    (SSDM2026)}, Long Beach, CA, June 8--10, 2026, Paper SSDM2026-183218.\\
    {\sffamily The published version of record can be found at:
    {\color{preprintlink}\mdseries\expandafter\url\expandafter{\vorlink}}}%
  }}}\par\vspace{8pt}%
}

\makeatletter
\patchcmd{\@maketitle}
  {\vspace*{-26pt}}
  {\vspace*{-26pt}\preprintbox}
  {}{\typeout{*** preprintbox patch FAILED ***}}
\makeatother
\confshortname{SSDM2026}
\conffullname{the ASME 2026\\ Aerospace Structures, Structural Dynamics, and Materials Conference}

\confdate{June 8-10}
\confyear{2026}
\confcity{Long Beach}
\confcountry{California}

\title{Modal Parameter Extraction via Propeller-Driven Vibration Testing}

\author{Gabriele Dessena\textsuperscript{1,}\thanks{Address all correspondence to this author. Email: {gdessena@ing.uc3m.es}.}$\;\;$\orcidlink{0000-0001-7394-9303} and Alessandro Pontillo\textsuperscript{2} \orcidlink{0000-0003-0015-685X}
    \affiliation{
	\textsuperscript{1}Department of Aerospace Engineering, Universidad Carlos III de Madrid, Leganés, Madrid, Spain\\\textsuperscript{2}School of Engineering, University of the West of England, Bristol, BS161QY, UK
    }	
}

\begin{document}

\maketitle    

%%%%%%%%%%%%%%%%%%%%%%%%%%%%%%%%%%%%%%%%%%%%%%%%%%%%%%%%%%%%%%%%%%%%%%
\begin{abstract}
{\it Ground Vibration Testing (GVT) supports aircraft certification but often requires lengthy and costly campaigns. Propeller-driven Vibration Testing (PVT) is assessed here as an output-only alternative, in line with Operational Modal Analysis approaches such as Taxi Vibration Testing and Flight Vibration Testing. A cantilever Aluminium 7075-T6 wing spar is instrumented with seven accelerometers and excited by an outboard electric motor and propeller. Seven runs are carried out: a motor-off baseline, five constant-throttle cases, and a manual up--down throttle sweep. The acquired spectra indicate that the dominant resonances remain observable under propeller excitation, while low-throttle conditions introduce narrowband harmonics that may mask structural peaks; the sweep reduces persistent overlap. Modal parameters are identified for the baseline and sweep cases using the Natural Excitation Technique with the Loewner Framework (NExT-LF). The first two modes remain closely matched (Modal Assurance Criterion (MAC) $>0.99$), whereas the third mode shows reduced repeatability (MAC $=0.827$) and a larger frequency shift, consistent with propeller-induced bending--torsion coupling and non-ideal sweep control. Overall, PVT provides a viable complement to GVT for extracting low-frequency modal information and motivates pursuing future work on automated throttle scheduling and coupling-aware test planning.}
\end{abstract}

%%%%%%%%%%%%%%%%%%%%%%%%%%%%%%%%%%%%%%%%%%%%%%%%%%%%%%%%%%%%%%%%%%%%%%
\begin{nomenclature}

% -------------------- Latin (symbols, abbreviations, acronyms) --------------------
\entry{3S}{Three-cell lithium polymer battery pack (three cells in series).}
\entry{ANPSD}{Average Normalised Power Spectral Density.}
\entry{AutoMAC}{Auto Modal Assurance Criterion.}
\entry{E}{Young's Modulus [Pa]}
\entry{EMA}{Experimental Modal Analysis.}
\entry{ESC}{Electronic speed controller.}
\entry{FEM}{Finite element model.}
\entry{$f_s$}{Sampling frequency [Hz].}
\entry{FVT}{Flight Vibration Testing.}
\entry{GVT}{Ground Vibration Testing.}
\entry{G}{Shear modulus [Pa].}
\entry{HAR}{High-aspect-ratio.}
\entry{$k$}{System order.}
\entry{$k_{\min}$}{Minimum system order used in stabilisation diagram construction.}
\entry{$k_{\max}$}{Maximum system order used in stabilisation diagram construction.}
\entry{$\mathbf{k}$}{Set of system orders used in stabilisation diagram construction.}
\entry{LF}{Loewner Framework.}
\entry{LiPo}{Lithium polymer (battery).}
\entry{MAC}{Modal Assurance Criterion.}
\entry{$\mathrm{MAC}_{\mathrm{stab}}$}{MAC stabilisation threshold for pole selection.}
\entry{NExT-LF}{Natural Excitation Technique with the Loewner Framework.}
\entry{OMA}{Operational Modal Analysis.}
\entry{PSD}{Power Spectral Density.}
\entry{PVT}{Propeller-driven Vibration Testing.}
\entry{RPM}{Revolutions per minute.}
\entry{SHM}{Structural Health Monitoring.}
\entry{$T_d$}{Acquisition duration [s].}
\entry{TVT}{Taxi Vibration Testing.}

% -------------------- Greek (symbols) --------------------
\entry{$\Delta k$}{System-order increment used in stabilisation diagram construction.}
\entry{$\Delta \omega_{\mathrm{stab}}$}{Frequency stabilisation tolerance for pole selection.}
\entry{$\Delta \zeta_{\mathrm{stab}}$}{Damping stabilisation tolerance for pole selection.}
\entry{$\zeta_n$}{Damping ratio of the $n$th mode.}
\entry{$\zeta_{\min}$}{Minimum admissible damping ratio for pole screening.}
\entry{$\zeta_{\max}$}{Maximum admissible damping ratio for pole screening.}
\entry{$\nu$}{Poisson's ratio.}
\entry{$\rho$}{Material density [kgm$^{-3}$].}
\entry{$\boldsymbol{\phi}_n$}{Mode shape of the $n$th mode.}
\entry{$\omega_n$}{Natural frequency of the $n$th mode [Hz].}
\entry{$\omega_{\min}$}{Lower bound of admissible natural-frequency range for pole screening [Hz].}
\entry{$\omega_{\max}$}{Upper bound of admissible natural-frequency range for pole screening [Hz].}

\end{nomenclature}
%%%%%%%%%%%%%%%%%%%%%%%%%%%%%%%%%%%%%%%%%%%%%%%%%%%%%%%%%%%%%%%%%%%%%%
\section*{INTRODUCTION}
Ground vibration testing (GVT) constitutes a fundamental step within the aircraft certification framework \cite{DeFlorio2011}. Consequently, GVT campaigns are typically extensive and technically demanding, often spanning several weeks and requiring complex experimental configurations \cite{Peeters2009a}. This inevitably results in substantial costs and significant time investment. Moreover, the adoption of advanced materials and innovative aircraft architectures has shown that the underlying assumptions of linearity are not always strictly valid \cite{lubrina2014}. Such nonlinear behaviour may originate from interactions between structural subcomponents, as shown for the Morane-Saulnier MS.760 Paris jet trainer in \cite{Kerschen2013}, and geometrical effects, as demonstrated in \cite{Dessena2022h} for a high-aspect-ratio (HAR) wing. Nonetheless, in the majority of practical situations, these effects can be neglected, since civil aircraft generally operate within a predominantly linear regime. Under these conditions, the primary outcomes of GVT remain the identification of modal parameters, namely natural frequencies ($\omega_n$), damping ratios ($\zeta_n$), and mode shapes ($\mathbf{\phi}_n$). Furthermore, vibration-based approaches relying on modal parameters have also been applied in Aeronautics and Aerospace for Structural Health Monitoring (SHM) for the detection and localisation of damage, most notably on the Space Shuttle vertical rudder \cite{Pappa1998}, and more recently in benchmark investigations on a flexible wing with simulated damage \cite{Dessena2022g}, as well as in vibration-based SHM assessments conducted on a damaged BAE Systems Hawk T1A full airframe \cite{Dessena2025a}. A detailed explanation of how vibration data, and in particular modal parameters, can be employed for this purpose is beyond the scope of this work; the interested reader is referred to the recent review in \cite{Farrar2025}. In general terms, structural damage may be interpreted as a deviation of the modal parameters with respect to the undamaged configuration.

For these reasons, several methodologies have been proposed to improve the efficiency of the GVT process. Among these, Taxi Vibration Testing (TVT) \cite{Govers2017} has attracted increasing attention over the past two decades. The core idea behind TVT is to exploit aircraft taxiing on a runway or taxiway, which is inherently uneven, as a natural source of excitation for dynamic testing. Within this framework, two main approaches can be identified: (i) towing-based excitation, as demonstrated in \cite{Pagani2021}, and (ii) active taxiing, where the excitation is provided by the aircraft propulsion system itself, with a demonstrator-scale application reported in \cite{Al-Bess2024}.

In parallel, output-only techniques have also enabled the extension of vibration testing to in-flight conditions through Flight Vibration Testing (FVT) \cite{Boeswald2017}, where modal parameters are identified directly from in-flight responses. Beyond fixed-wing aircraft, such approaches have found relevant aerospace applications in the identification of launch vehicles, where operational modal analysis has been successfully applied to flight data to track modal parameters under time-varying conditions associated with propellant consumption \cite{Eugeni2018}.

From a methodological standpoint, these approaches differ from classical Experimental Modal Analysis (EMA), on which traditional GVT is based, where the excitation is known, measured, and intentionally applied to the structure \cite{Rainieri2014}. In contrast, both TVT and FVT are typically formulated within the framework of Operational Modal Analysis (OMA), which relies exclusively on response measurements and assumes the excitation to be unknown and broadband in nature.

Building upon the use of OMA in Aeronautics, this work considers Propeller-driven Vibration Testing (PVT), as recently introduced in \cite{Dessena2025b}. The underlying premise is that, in the context of the growing adoption of electric and hybrid-electric regional aircraft, PVT may provide an additional means for the dynamic validation of aircraft structures. In this sense, PVT has the potential to reduce reliance on conventional Ground Vibration Testing (GVT), alleviating associated time and cost constraints, while in the longer term complementing TVT and FVT. From a technical standpoint, PVT can exploit the broadband forcing generated by the propeller action and, potentially, its rotational effects to excite the wing.

The primary objective of this study is to demonstrate the feasibility of extracting modal parameters from PVT data and to compare the identified results with those obtained from a classical impact hammer test. To this end, the concept of PVT is first recalled and subsequently applied to a flexible wing spar model derived from the Experimental Flight Dynamics Testing for Highly Flexible Aircraft project at the University of Bristol \cite{BannehekaNavaratna2022}. The experimental campaign is structured as follows: an initial test is conducted with the propeller inactive to establish a baseline reference; this is followed by a set of tests performed at increasing throttle settings (and so propeller speed), corresponding to different propeller rotational speeds; finally, a continuous voltage sweep is executed.

The subsequent sections describe the experimental setup and the tested specimen, after which the identified modal parameters are presented and discussed.

\section*{\uppercase{Materials and Methods}}
\subsection*{{Flexible Wing Spar}}
The wing spar considered in this work spans a total free length of 80 cm, measured from the end of the constrained region, which extends over 7 cm. The spar has a constant thickness of 2.3 mm along its entire span. The specimen is manufactured from Aluminium 7075-T6, and is therefore characterised by standard aluminium mechanical properties in terms of density ($\rho =$ 2810 kgm\textsuperscript{-3}), Young’s modulus (E = 71.7 GPa), shear modulus (G = 26.9 GPa) and Poisson’s ratio ($\nu=$ 0.33).

\begin{figure*}[!ht]
\centering
\includegraphics[width=0.7\textwidth]{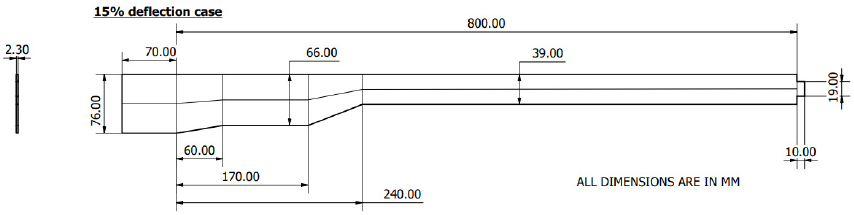}
\caption{\uppercase{Planform geometry of the flexible wing spar (retrieved from }\cite{BannehekaNavaratna2022}).}
\label{fig:fig1} 
\end{figure*}

As shown in \cref{fig:fig1}, the spar exhibits a non-prismatic geometry composed of four sections with varying cross-sectional height. From the constrained end, an initial region with maximum depth is followed by a stepped transition leading to a reduced-height section. A second geometric transition then connects to the final, slender portion of the spar, which extends over the majority of the span and maintains a constant minimum height.

This stepped and tapered configuration results in a gradual reduction of stiffness along the span, representative of lightweight wing primary structures, while preserving a level of geometric simplicity suitable for experimental investigation.
\subsection*{{Experimental Setup}}
The constrained root of the spar is rigidly secured to a wall-mounted fixture, resulting in a cantilever beam-like boundary condition. Along the span, an in-house designed and manufactured motor bracket is installed at a distance of 32 cm from the spar tip. The bracket is produced via additive manufacturing using carbon-fibre-reinforced nylon filament and serves as the support for a TMOTOR AS2317 electric motor equipped with a two-blade, 8-inch propeller. The motor is driven by a Favourite FVT 50A BLHeli\_32 electronic speed controller, powered by a 3S lithium polymer battery, and interfaced with a receiver connected to a radio controller, which is used to control the electric motor.

\begin{figure*}[!ht]
\centering
\includegraphics[width=.8\textwidth]{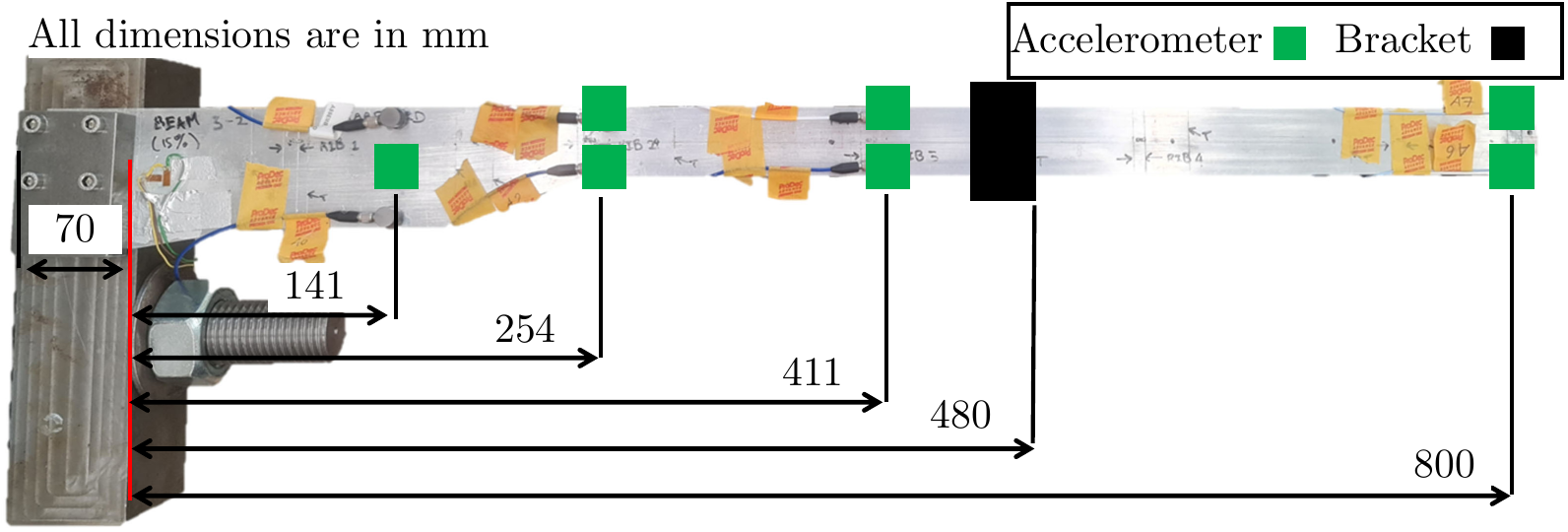}
\caption{\uppercase{Experimental setup. Apparent geometric distortions in the image are attributable solely to camera lens effects. (retrieved from }\cite{Dessena2025b}).}
\label{fig:fig2} 
\end{figure*}

The spar–bracket–motor assembly is instrumented with seven mono-axial accelerometers distributed along the span, as shown in \cref{fig:fig2}. Five sensors have a nominal sensitivity of approximately 100 mVg\textsuperscript{-1}, while the two accelerometers located at the spar tip feature a lower sensitivity of 10 mVg\textsuperscript{-1} to accommodate the higher expected vibration levels. The accelerometer positions were determined based on a preliminary finite element model of the spar. In particular, the sensor layout was selected by minimising the sum of the off-diagonal terms in the upper triangular portion of the Auto Modal Assurance Criterion (AutoMAC) matrix, to reduce modal correlation and facilitate independent identification of adjacent targeted modes.

The accelerometer signals are acquired using a National Instruments data acquisition board, controlled from MATLAB via a custom in-house script. All measurements are carried out on the same spar--bracket--motor--propeller configuration introduced in the previous section, which is therefore treated as the baseline setup.

Seven vibration tests are performed, as reported in \cref{tab:tab1}. A reference case is first acquired with the propeller stationary and the motor switched off (case i). The remaining steady-state tests are conducted at five constant throttle settings, namely 25, 37.5, 50, 62.5, and 75\% (cases ii--vi, respectively). Finally, an additional run is carried out by varying the throttle in an up-and-down sweep between 12.5\% and 77.5\% (case vii). Owing to the lack of hardware for automated throttle scheduling at the time of testing, the sweep is executed manually, using a stopwatch as a timing reference to approximate a linear throttle variation.

\begin{table}[!ht]
\caption{\uppercase{Summary of vibration test cases performed on the baseline spar configuration.}\label{tab:tab1}}
\centering
\begin{tabular}{c c c c}
\hline
Case & Motor condition & Throttle setting & $\sim$Duration $T_d$ [s] \\
\hline
i   & Motor off        & --               & $120$ \\
ii  & Motor on         & 25\%             & $300$ \\
iii & Motor on         & 37.5\%           & $300$ \\
iv  & Motor on         & 50\%             & $300$ \\
v   & Motor on         & 62.5\%           & $300$ \\
vi  & Motor on         & 75\%             & $300$ \\
vii & Motor on (sweep) & 12.5--77.5\%     & $600$ \\
\hline
\end{tabular}
\end{table}

For the motor-off baseline (case i), excitation is provided by two separate hammer impacts, with the second hit applied only once the response from the first has fully decayed. The input force is not measured; therefore, the dataset is treated as output-only. Consistently, modal identification for all cases, including case i, is performed within an Operational Modal Analysis (OMA) setting using the Natural Excitation Technique with the Loewner Framework (NExT-LF) \cite{Dessena2024f}. In all tests, only the seven accelerometer outputs shown in \cref{fig:fig2} are recorded. The acquisition times are approximately $T_d \approx 120$~s for case i, $T_d \approx 300$~s for cases ii--vi, and $T_d \approx 600$~s for case vii. All channels are sampled at $f_s = 1066$~Hz.

\section*{\uppercase{Results}}
\begin{figure}[htp!]
\centering
\includegraphics[width=\columnwidth]{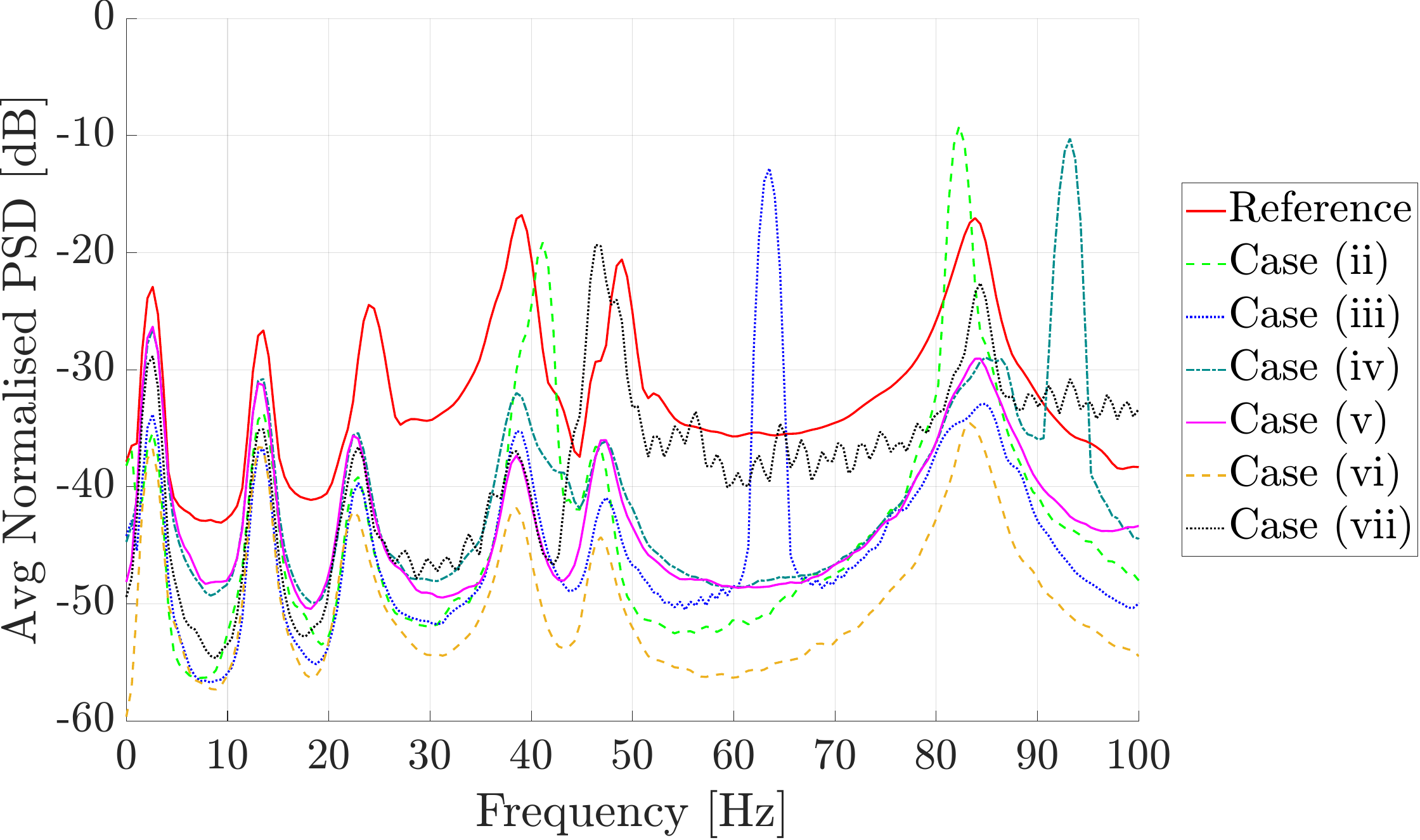}
\caption{\uppercase{ANPSD comparison across all test cases (impulse baseline and PVT). Adapted from }\cite{Dessena2025b}.}
\label{fig:fig3}
\end{figure}

The results from the seven tests are reported in \cref{fig:fig3}, where the Average Normalised Power Spectral Density (ANPSD) is shown for each operating condition. For each test, the PSDs of the seven accelerometer signals are normalised and then averaged, so that the comparison emphasises frequency content and resonant features rather than channel-dependent amplitude differences. This ANPSD representation is adopted as an initial output-only screening step, providing a compact view of the frequency bands effectively excited under each condition and of the tonal components introduced by propeller rotation. Modal parameters are subsequently identified only for cases (i) and (vii) using NExT-LF.

The motor-off impulse test (case i), with the propeller stationary (reference, red curve in \cref{fig:fig3}), is used as a baseline. Across all operating conditions, the ANPSD exhibits peaks that remain aligned with those of the reference case, indicating that the same structural modes are consistently activated. This provides an initial indication that propeller-induced excitation can reveal the relevant low-frequency dynamics of the wing spar, supporting the use of PVT as an output-only alternative in this context.

At the same time, \cref{fig:fig3} highlights a practical limitation at lower throttle settings. For throttle levels up to 50\% (cases ii--iv), the response includes pronounced narrowband components associated with propeller harmonics, visible at approximately 42, 63, 84, and 94~Hz. These tones may locally mask structural peaks and hinder identification if overlap occurs. By contrast, for higher throttle settings (cases v and vi), the dominant harmonic content is shifted above 100~Hz and does not interfere with the frequency range analysed here.

Since the location of propeller harmonics is not always available \emph{a priori}, case vii adopts an up--down throttle sweep to mitigate persistent harmonic overlap by spreading the tonal content over a broader frequency interval. The ANPSD confirms that this strategy reduces the prominence of discrete harmonic lines within the analysed range, albeit at the cost of a noisier spectral estimate compared with fixed-throttle tests. This is reasonably attributed to the manual execution of the sweep, timed using a stopwatch, due to the lack of dedicated hardware for automated throttle scheduling at the time of testing.

\begin{figure}[!ht]
\centering
	\begin{subfigure}[t]{\columnwidth}
	\centering
		{\includegraphics[width=\textwidth,keepaspectratio]{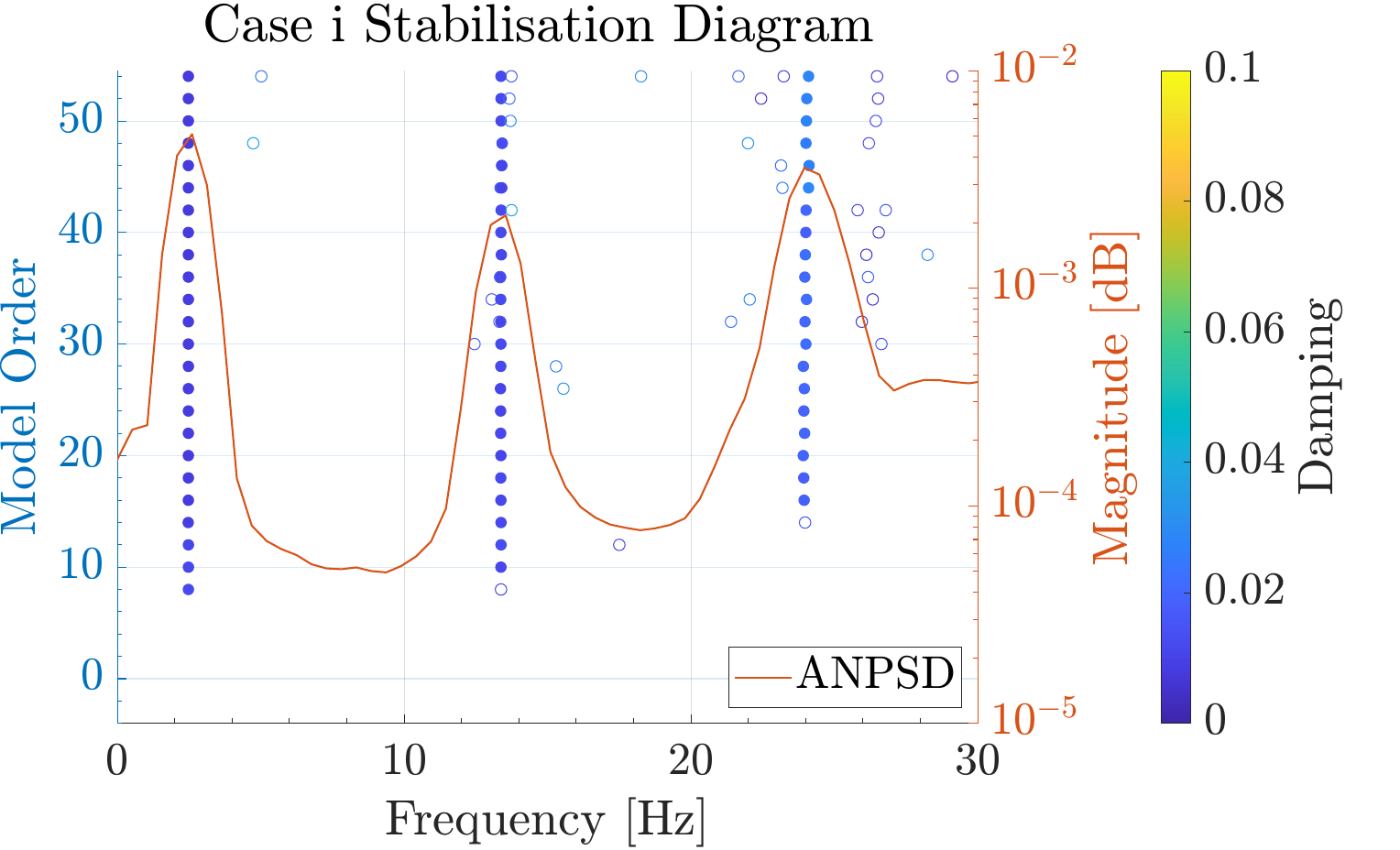}}
		\captionsetup{font={it},justification=centering}
		\subcaption{\label{fig:fig4a}}	
	\end{subfigure}
    \begin{subfigure}[t]{\columnwidth}
	\centering
		\includegraphics[width=\textwidth,keepaspectratio]{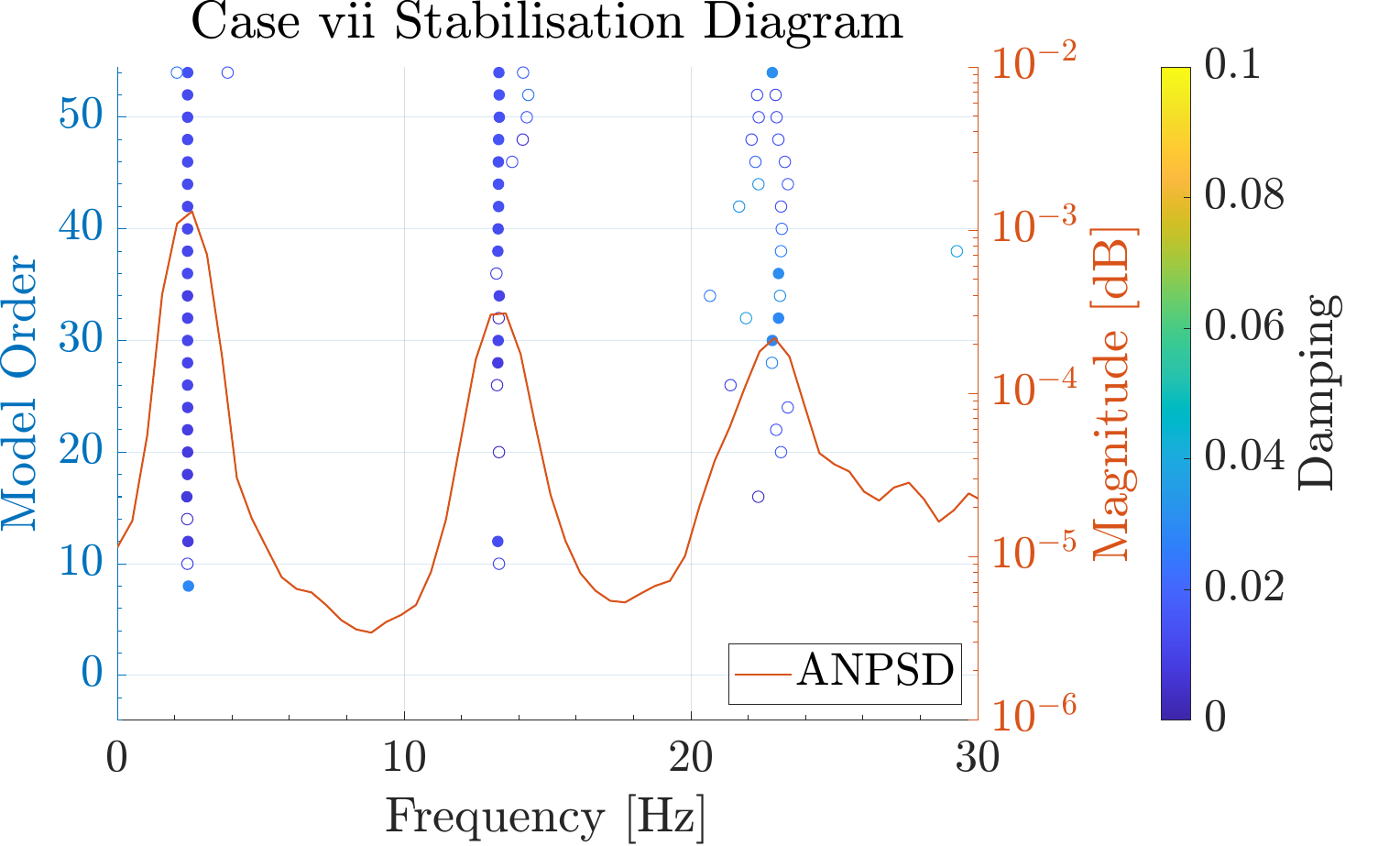}
		\captionsetup{font={it},justification=centering}
		\subcaption{\label{fig:fig4b}}	
	\end{subfigure}
	\caption{\uppercase{Stabilisation diagrams obtained via NExT-LF for cases (i) and (vii).}}
	\label{fig:fig4}
\end{figure}

Modal identification is introduced by inspecting the stabilisation diagrams obtained via NExT-LF for the motor-off reference case i and the sweep case vii, shown in \cref{fig:fig4a,fig:fig4b}. The diagrams are constructed by varying the system order from $k_{\min}=32$ to $k_{\max}=60$ with increment $\Delta k = 2$, yielding the extraction set $\mathbf{k}=\{\,k_{\min}:\Delta k:k_{\max}\,\}$. Hard screening is imposed through the admissible damping range $\zeta_n \in [\zeta_{\min}, \zeta_{\max}] = [0.005,\,0.03]$ and the natural frequency range $\omega_n \in [\omega_{\min}, \omega_{\max}] = [0,\,30]$. Stable poles are then selected using soft criteria based on frequency stabilisation $\Delta \omega_{\mathrm{stab}} = 0.01$, damping stabilisation $\Delta \zeta_{\mathrm{stab}} = 0.05$, and mode-shape consistency enforced through $\mathrm{MAC}_{\mathrm{stab}} = 0.95$.

In both cases, the stabilised poles form clear vertical clusters that align with the dominant ANPSD peaks, providing a consistent basis for modal selection. Nevertheless, the highest-frequency mode within the analysed set (3\textsuperscript{rd} mode, $\sim 25$~Hz) is less robustly stabilised in case vii, with increased scatter across model order. This behaviour is consistent with a stronger contribution of bending--torsion coupling under propeller operation, where the active propeller introduces an additional torsional moment. The effect is plausibly exacerbated by the non-ideal character of the manual sweep, which introduces variability in the excitation rate and departs from strictly stationary conditions.

\begin{figure*}[!ht]
\centering
	\begin{subfigure}[t]{\columnwidth}
	\centering
		{\includegraphics[width=\textwidth,keepaspectratio]{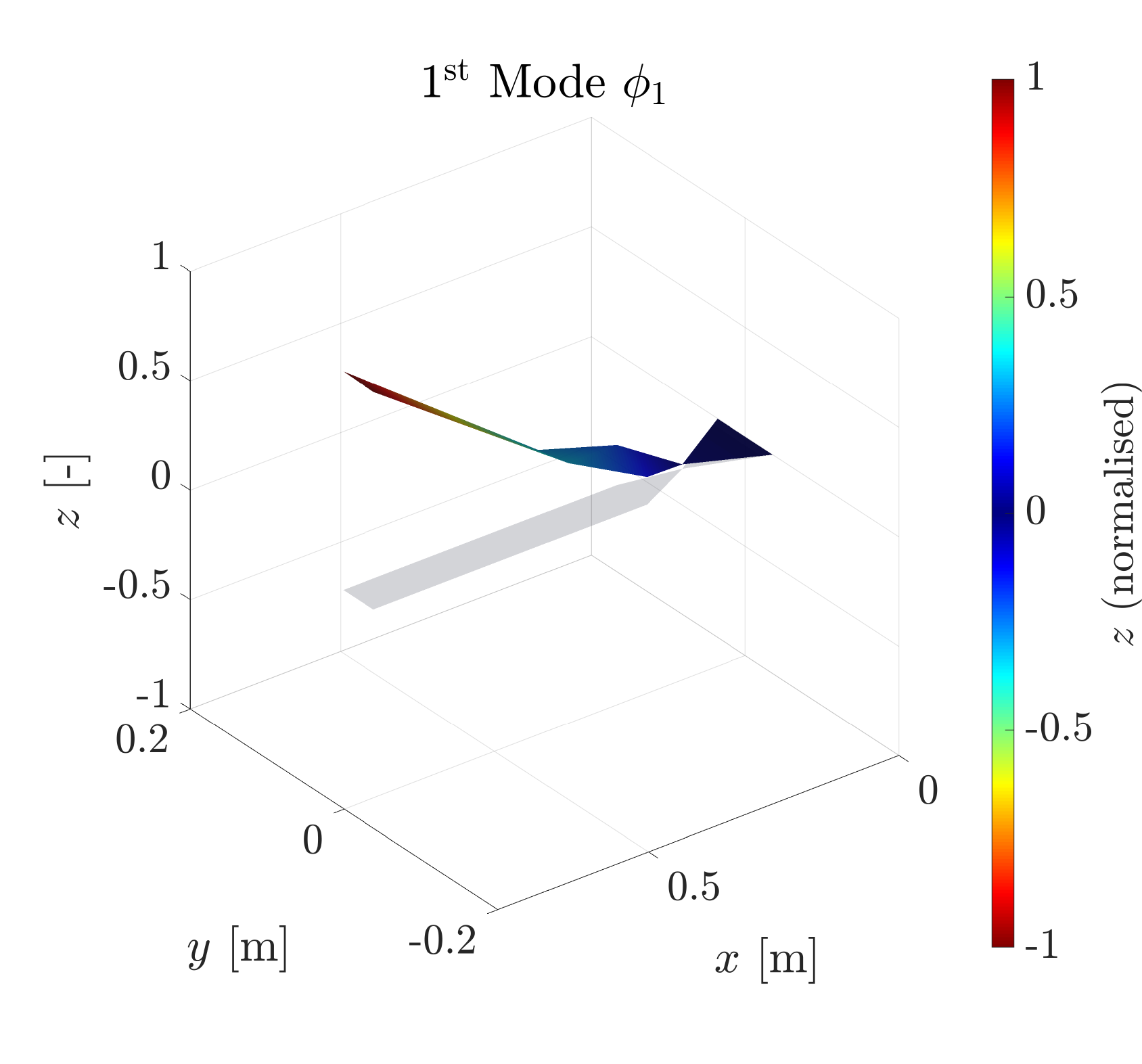}}
		\captionsetup{font={it},justification=centering}
		\subcaption{\label{fig:fig5a}}	
	\end{subfigure}
    \begin{subfigure}[t]{\columnwidth}
	\centering
		\includegraphics[width=\textwidth,keepaspectratio]{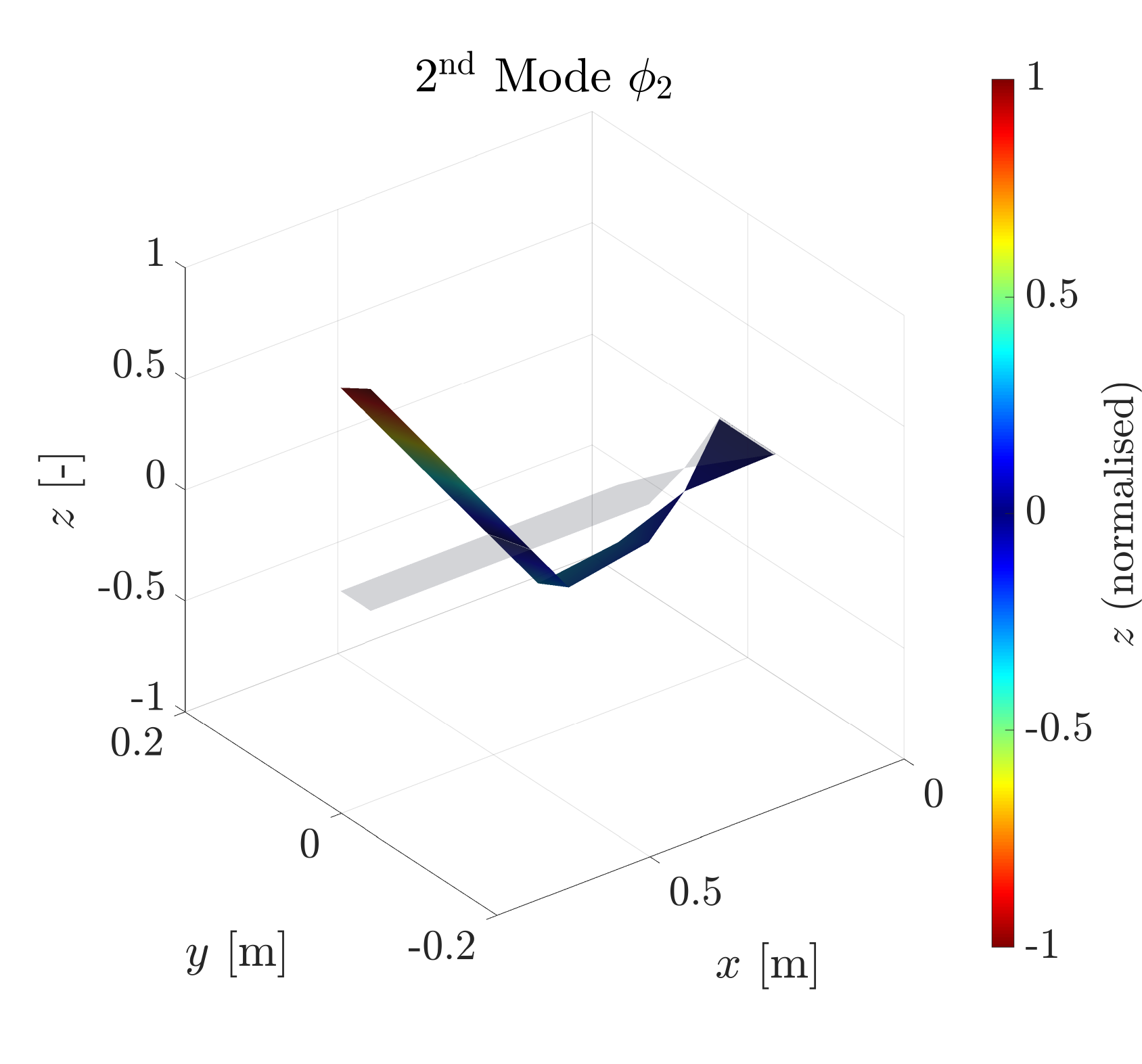}
		\captionsetup{font={it},justification=centering}
		\subcaption{\label{fig:fig5b}}	
	\end{subfigure}
        \begin{subfigure}[t]{\columnwidth}
	\centering
		\includegraphics[width=\textwidth,keepaspectratio]{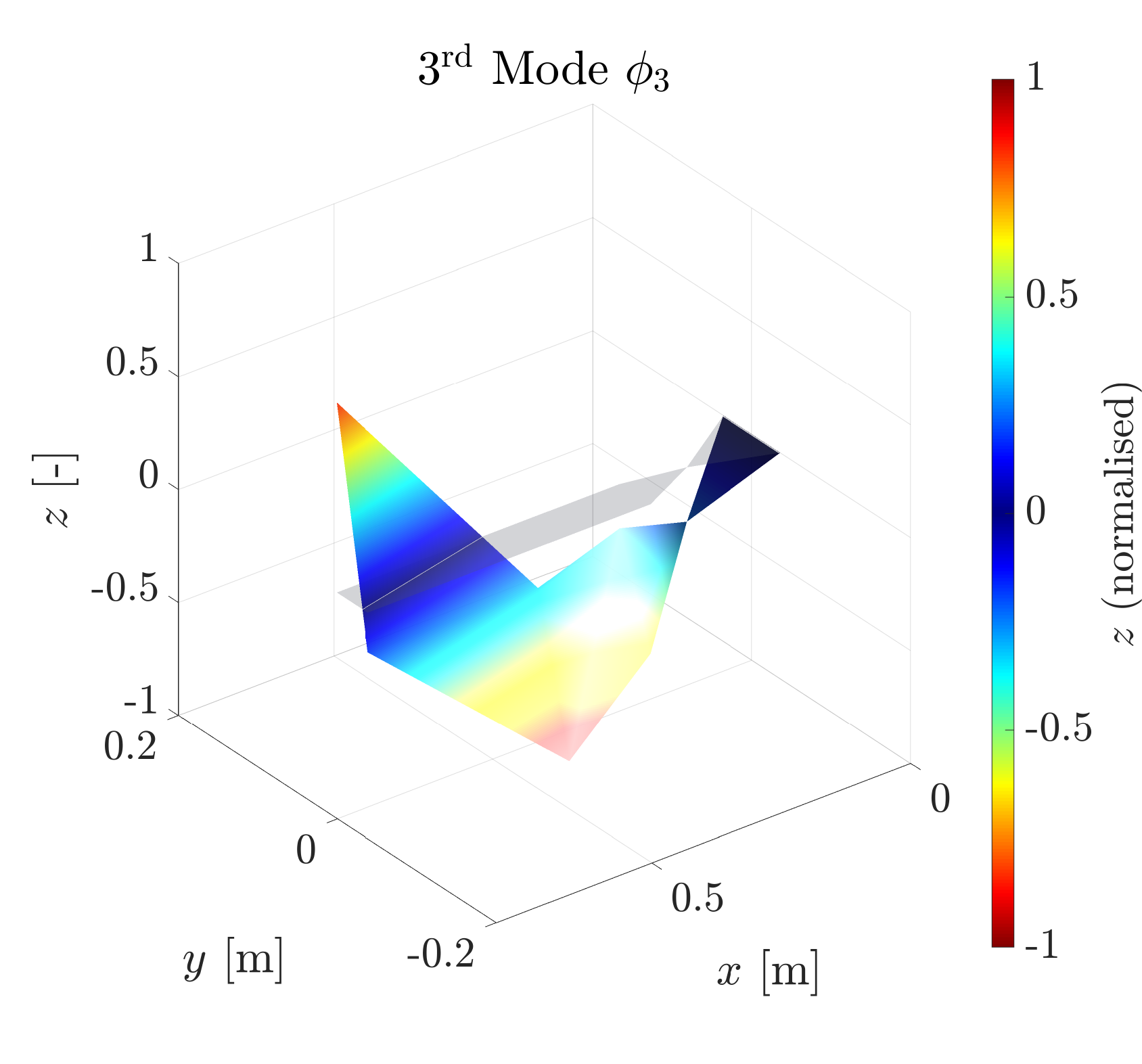}
		\captionsetup{font={it},justification=centering}
		\subcaption{\label{fig:fig5c}}	
	\end{subfigure}
	\caption{\uppercase{Mode shapes $\boldsymbol{\phi}_{1-3}$ identified via NExT-LF for case (vii).}}
	\label{fig:fig5}
\end{figure*}

Given this premise, the identified modal parameters are presented for the reference configuration (case~i) and for the PVT sweep condition (case~vii). The natural frequencies and damping ratios are summarised in \cref{tab:tab2,tab:tab3}, while the corresponding mode shapes are reported in \cref{fig:fig5a,fig:fig5b,fig:fig5c}. Mode-shape agreement is quantified through the MAC with respect to the reference, yielding $\mathrm{MAC}_1 = 0.999$, $\mathrm{MAC}_2 = 0.998$, and $\mathrm{MAC}_3 = 0.827$.

\begin{table}[!htp]
\caption{\uppercase{Natural frequencies identified for the reference case and case~(vii).}\label{tab:tab2}}
\centering
\begin{tabular}{c c c}
\hline
Mode \# & Reference [Hz] & Case vii [Hz] -- (difference [\%]) \\
\hline
$1$ & $2.48$  & $2.45$\\
& & ($-1.21$) \\
$2$ & $13.37$ & $13.30$\\ 
& &($-0.52$) \\
$3$ & $24.10$ & $22.84$\\ 
& & ($-5.23$) \\
\hline
\end{tabular}
\end{table}

As shown in \cref{tab:tab2}, the first two natural frequencies change only marginally between cases i and vii. By contrast, the third frequency decreases by $5.23\%$, in line with the weaker stabilisation observed for the last mode under sweep-driven excitation.

\begin{table}[!ht]
\caption{\uppercase{Damping ratios identified for the reference case and case~(vii).}\label{tab:tab3}}
\centering
\begin{tabular}{c c c}
\hline
Mode \# & Reference [-] & Case vii [-] -- (difference [\%]) \\
\hline
$1$ & $0.008$ & $0.014$\\
& &($+75.00$) \\
$2$ & $0.012$ & $0.015$\\
& & ($+25.00$) \\
$3$ & $0.028$ & $0.031$\\
& & ($+10.71$) \\
\hline
\end{tabular}
\end{table}

Regarding damping, \cref{tab:tab3} shows an increase for all modes in case vii, with the largest relative change for the first mode. This trend is compatible with additional dissipation mechanisms activated under propeller operation, together with the non-ideal sweep conditions, which can bias damping estimates when excitation properties vary in time.

The mode shapes identified for the PVT sweep case (vii) are shown in \cref{fig:fig5}. The first and second modes (\cref{fig:fig5a,fig:fig5b}) remain consistent with the reference configuration, as also reflected by $\mathrm{MAC}_1 = 0.999$ and $\mathrm{MAC}_2 = 0.998$. A different trend is observed for the third mode (\cref{fig:fig5c}), where a visible deviation is present and the agreement decreases to $\mathrm{MAC}_3 = 0.827$. This value nevertheless remains within commonly adopted thresholds for correlation, and the third mode can therefore still be classified as correlated, albeit with reduced similarity compared with the first two modes. The discrepancy is consistent with the reduced stabilisation and larger frequency shift observed for this mode, and is plausibly associated with increased sensitivity to torsional contributions under propeller-induced excitation, compounded by the manual sweep.

Overall, the results indicate that PVT can recover a consistent set of modal parameters for the flexible wing spar, with excellent agreement for the first two modes and acceptable correlation for the third. These findings provide initial evidence that propeller-induced excitation can be exploited for OMA-based modal testing of flexible wings, while suggesting that improved sweep control and further mitigation of torsional contamination would be beneficial for higher-order modes.

\section*{\uppercase{Conclusions and future work}}
This work investigates \emph{Propeller-driven Vibration Testing} (PVT) as an output-only approach for modal parameter extraction on a cantilever wing spar equipped with an outboard electric motor and propeller. The comparison of the {Average Normalised Power Spectral Density} (ANPSD) across the tested conditions shows that the dominant resonant peaks observed in the motor-off baseline remain visible under propeller excitation, indicating that the same structural modes are observable within the frequency range of interest. At low throttle, the response is affected by narrowband propeller harmonics, which may locally mask structural resonances and hinder identification when overlap occurs. For this reason, an up--down throttle sweep is adopted to reduce persistent harmonic overlap within the analysed band.

Modal identification is carried out using the {Natural Excitation Technique with the Loewner Framework} (NExT-LF) for the reference case i and the sweep case vii. The stabilisation diagrams show consistent pole clusters across model order, supporting mode selection based on persistent solutions. Excellent agreement is obtained for the first two modes under PVT, with negligible frequency variations and {Modal Assurance Criterion} (MAC) values close to unity. The third mode exhibits reduced stabilisation and lower shape agreement, while remaining within commonly adopted thresholds used to classify mode shapes as correlated. This behaviour is consistent with an increased sensitivity of the higher-frequency mode to bending--torsion interaction, which becomes more relevant under propeller operation.

Future work focuses on improving repeatability and explicitly anticipating propeller-induced coupling effects prior to testing and identification. In particular:
\begin{itemize}
    \item \textit{Automation of the sweep:} implement automated throttle scheduling to enforce a prescribed sweep law and reduce variability introduced by manual operation. A practical route is closed-loop control based on measured revolutions per minute, using {Electronic Speed Controllers} telemetry or an external sensor, enabling repeatable sweeps and dwell tests targeted at specific frequency bands.
    \item \textit{Pre-test treatment of bending--torsion coupling:} develop a predictive framework to account for the additional torsional moment and gyroscopic contributions induced by the rotating propeller. A reduced-order structural model, augmented with a propulsion-induced moment description, can be used to identify coupling-sensitive modes beforehand, guide sensor placement, and interpret shifts and scatter in the higher-frequency pole estimates.
    \item \textit{Extension and validation:} apply the methodology to more representative wing assemblies and, subsequently, aircraft-relevant configurations, benchmarking PVT-derived parameters against established test practices under comparable boundary conditions.
\end{itemize}

Overall, the results support PVT as a viable complement to {Operational Modal Analysis} (OMA)-based ground-test capability, while indicating that improved excitation control and an explicit consideration of propeller-induced bending--torsion coupling are key to strengthening the method reliability for higher-order modes identification.

\bibliographystyle{asmems4}

%%%%%%%%%%%%%%%%%%%%%%%%%%%%%%%%%%%%%%%%%%%%%%%%%%%%%%%%%%%%%%%%%%%%%%
\begin{acknowledgment}
The authors thank Prof Mark H. Lowenberg from the School of Civil, Aerospace and Design Engineering at the University of Bristol for kindly allowing the use of the wing spar, which was designed and manufactured within the EPSRC Experimental Flight Dynamics Testing for Highly Flexible Aircraft project grant ref. EP/T018739/1. 
This work has been supported by the Madrid Government (\textit{Comunidad de Madrid}, Spain) under the Multiannual Agreement with the UC3M (IA\_aCTRl-CM-UC3M).
\end{acknowledgment}

%%%%%%%%%%%%%%%%%%%%%%%%%%%%%%%%%%%%%%%%%%%%%%%%%%%%%%%%%%%%%%%%%%%%%%
% The bibliography is stored in an external database file
% in the BibTeX format (file_name.bib).  The bibliography is
% created by the following command and it will appear in this
% position in the document. You may, of course, create your
% own bibliography by using thebibliography environment as in
%
% \begin{thebibliography}{12}
% ...
% \bibitem{itemreference} D. E. Knudsen.
% {\em 1966 World Bnus Almanac.}
% {Permafrost Press, Novosibirsk.}
% ...
% \end{thebibliography}

% Here's where you specify the bibliography database file.
% The full file name of the bibliography database for this
% article is asme2e.bib. The name for your database is up
% to you.
%\bibliography{ASME}

%%%%%%%%%%%%%%%%%%%%%%%%%%%%%%%%%%%%%%%%%%%%%%%%%%%%%%%%%%%%%%%%%%%%%%
% \appendix       %%% starting appendix
% \section*{Appendix A: Head of First Appendix}
% Avoid Appendices if possible.

% %%%%%%%%%%%%%%%%%%%%%%%%%%%%%%%%%%%%%%%%%%%%%%%%%%%%%%%%%%%%%%%%%%%%%%
% \section*{Appendix B: Head of Second Appendix}
% \subsection*{Subsection head in appendix}
% The equation counter is not reset in an appendix and the numbers will
% follow one continual sequence from the beginning of the article to the very end as shown in the following example.
% \begin{equation}
% a = b + c.
% \end{equation}

\end{document}